\documentclass[sigconf]{acmart}

\AtBeginDocument{%
  \providecommand\BibTeX{{%
    \normalfont B\kern-0.5em{\scshape i\kern-0.25em b}\kern-0.8em\TeX}}}
    
\usepackage{soul}
\usepackage{balance}
\usepackage{ifthen}
\usepackage{amssymb}
\newboolean{showcomments}
\setboolean{showcomments}{true} % toggle to show or hide comments
\ifthenelse{\boolean{showcomments}}
  {\newcommand{\nb}[2]{
    \fcolorbox{gray}{yellow}{\bfseries\sffamily\scriptsize#1}
    {\sf\small$\blacktriangleright$\textit{#2}$\blacktriangleleft$}
   }
   
  }
  {\newcommand{\nb}[2]{}
   
  }

\copyrightyear{2020}
\acmYear{2020}
\setcopyright{acmlicensed}
\acmConference[MSR '20]{17th International Conference on Mining Software Repositories}{October 5--6, 2020}{Seoul, Republic of Korea}
\acmBooktitle{17th International Conference on Mining Software Repositories (MSR '20), October 5--6, 2020, Seoul, Republic of Korea}
\acmPrice{15.00}
\acmDOI{10.1145/3379597.3387500}
\acmISBN{978-1-4503-7517-7/20/05}

\begin{document}

%%
%% The "title" command has an optional parameter,
%% allowing the author to define a "short title" to be used in page headers.
\title[Dataset for Identity Resolution]{A Dataset and an Approach for Identity Resolution of 38 Million Author IDs extracted from 2B Git Commits}

%%
%% The "author" command and its associated commands are used to define
%% the authors and their affiliations.
%% Of note is the shared affiliation of the first two authors, and the
%% "authornote" and "authornotemark" commands
%% used to denote shared contribution to the research.
% \numberofauthors{4}
\author{Tanner Fry}
\email{tfry2@vols.utk.edu}
\author{Tapajit Dey}
\email{tdey2@vols.utk.edu}
\affiliation{%
%   \department{Electrical Engineering and Computer Science}
  \institution{The University of Tennessee}
  \streetaddress{1520 Middle Dr.}
  \city{Knoxville}
  \state{TN}
  \country{USA}
  \postcode{37996}
}
\author{Andrey Karnauch}
\email{akarnauc@vols.utk.edu}

\author{Audris Mockus}
\email{audris@mockus.org}
\affiliation{%
%   \department{Electrical Engineering and Computer Science}
  \institution{The University of Tennessee}
  \streetaddress{1520 Middle Dr.}
  \city{Knoxville}
  \state{TN}
  \country{USA}
  \postcode{37996}
}

%%
%% By default, the full list of authors will be used in the page
%% headers. Often, this list is too long, and will overlap
%% other information printed in the page headers. This command allows
%% the author to define a more concise list
%% of authors' names for this purpose.
%\renewcommand{\shortauthors}{Trovato and Tobin, et al.}

%%
%% The abstract is a short summary of the work to be presented in the
%% article.
\begin{abstract}
The data collected from open source projects provide means to model large software ecosystems, but often suffer from data quality issues, specifically, multiple author identification strings in code commits might actually be  associated with one developer. While many methods have been proposed for addressing this problem, they are either heuristics requiring manual tweaking, or require too much calculation time to do pairwise comparisons for 38M author IDs in, for example, the World of Code collection. In this paper, we propose a method that finds all author IDs belonging to a single developer in this entire dataset, and share the list of all author IDs that were found to have aliases. To do this, we first create blocks of potentially connected author IDs and then use a machine learning model to predict which of these potentially related  IDs belong to the same developer. We processed around 38 million author IDs and found around 14.8 million IDs to have an alias, which belong to 5.4 million different developers, with the median number of aliases being 2 per developer. This dataset can be used to create more accurate models of developer behaviour at the entire OSS ecosystem level and can be used to provide a service to rapidly resolve new author IDs.
\end{abstract}

%%
%% The code below is generated by the tool at http://dl.acm.org/ccs.cfm.
%% Please copy and paste the code instead of the example below.
%% Keywords. The author(s) should pick words that accurately describe
%% the work being presented. Separate the keywords with commas.

%% A "teaser" image appears between the author and affiliation
%% information and the body of the document, and typically spans the
%% page.

%%
%% This command processes the author and affiliation and title
%% information and builds the first part of the formatted document.
\maketitle

\section{Introduction}\label{s:intro}

The studies of open source software ecosystems often rely on investigating the digital traces left by the software developers, for the reconstruction and quantification of the behavior of an individual~\cite{dey2019patterns,amreen2019methodology}, a team~\cite{10.1145/337180.337209}, or an organization~\cite{10.1145/1882291.1882311,Hackbarth2010}. However, the data obtained directly by mining software repositories often suffer from quality issues. In this paper, we focus on the
challenge of identifying authorship of code changes (commits) based on the information that can be obtained from the 
multitude of open source repositories, which is a deceptively hard problem to address. Software developers often use multiple email addresses, different variations of their names, or aliases while creating commits, and being able to identify the different IDs used by an author is crucial for improving the quality of the data obtained through mining different software repositories.

A number of ways to address this problem are available in the
literature, but they are either based on very simple heuristics with
little room for tweaking, e.g. from simple name matching with manual
inspection~\cite{german2003automating}, to very complex techniques 
that require full pairwise comparisons~\cite{amreen2019alfaa}. In our case of
38M author IDs obtained from the World of Code~\cite{woc19}
dataset, the full pairwise comparison involves
$10^{14}$ computationally expensive comparisons, which  
is beyond the reach of the fastest super computers.
In this paper, we
are proposing an approach for author identity resolution that
requires nothing more than the first name, last name, and email
address of a commit author, but is trained with a Random Forest
model for improved accuracy. We start by
first dividing all the author IDs into blocks of authors IDs using a
combination of heuristic methods to ensure 
a high probability of in-block linkage and a low probability of
out-of-block linkage between the IDs. To make the computation
possible, we assume that
author IDs belonging to different blocks are not linked, and use
pairwise comparison followed by transitive closure to identify the
author IDs belonging to a block that are linked with each other.

We applied our approach on 38,362,013 author IDs collected from the
World of Code data~\cite{woc19}. Using a combination of three heuristics,
we found 15,177,184 author IDs in 5,508,119 blocks of two or
more IDs per block, with the remaining author IDs being in blocks of size
1. Finally, after applying pairwise comparison, followed by transitive closure, we found that 14,861,538 IDs had an alias, which were found to be
associated with 5,427,024 different developers. We tested the
accuracy of our method by enlisting the help of 44 developers, who
identified 207 different IDs that belonged to them. In terms of
identifying pairs of author IDs linked together, our method achieved
in a precision of 0.99 and a recall of 0.84 using this test data. 

We created a dataset with all the author IDs that were found to have
an alias (alternate ID), which is a compressed CSV file with `;' as the
separator. If an author was found to have 2 different IDs:
\textit{I1, I2}, then it is recorded in the file in 2 separate
lines, with the lines being \texttt{I1;I1} and \texttt{I1;I2},
i.e. the first column is the group identifier, which is one of the
IDs in a group, and the second column contains the different author
IDs in separate lines. 

However, sharing this dataset publicly has some ethical implications.
Although the data is collected from public git commits to public repositories, 
we can not put the list of emails in a plain text form as any email
harvester may be able to obtain the data.  To enable replication and the
ability to use data in MSR studies, we release all data with restricted access,
so that anyone who adequately  certifies that
they will only use the data in a research study and not for commercial
purpose can get access to the data. The data with
restricted access is available at:
\textcolor{red}{\url{https://zenodo.org/record/3653283}}.
We also provide completely publicly the SHA1 values of author ID map
(SHA1 to all related SHA1s) so anyone who already has that author ID can
easily calculate SHA1 and use our data for linking it to other author IDs. That way
the MSR studies can be enabled without the need to expose the personal IDs.
The publicly available version of the data is available at:
\textcolor{red}{\url{https://zenodo.org/record/3700859}}.
Moreover, we suggest researchers using our data familiarize themselves with 
the latest data confidentiality laws and be aware of the related privacy concerns 
while using the data. 

We are also sharing the dataset with the author blocks, along with the block id,
frequency of the first name (number of times it appears in the 38M WoC version Q
author IDs), frequency of the last name, full name, email, and the Author ID.
This is the result of the first stage of processing, and can be used by the
community for using different types of pairwise comparisons for further refining
the final result. The restricted access version of the data (because of similar
reasons) is available at:
\textcolor{red}{\url{https://zenodo.org/record/3648702}}, and the publicly
available version of the data (with email addresses and author IDs of 
individual authors replaced by their corresponding hash values) is available at:
\textcolor{red}{\url{https://zenodo.org/record/3701819}}.

We also share the model used to predict if the two IDs are identical
as well as the additional blocking of author IDs using a combination
of three heuristics (see Section~\ref{s:link}). Anyone using the source code we have provided can
run the model on these larger blocks, thus potentially increasing the 
accuracy of the result. The scripts and the model are available at: 
\textcolor{red}{\url{https://zenodo.org/record/3653069}}.

The rest of the paper is organized as follows: In
Section~\ref{s:sources}, we describe the data source. Our approach
for linking author identities is described in
Section~\ref{s:link}. We describe our plans to further refine our
method in Section~\ref{s:future}. The limitations to the approach is
described in Section~\ref{s:lim}, and we conclude the paper in
Section~\ref{s:conc}. 

\vspace{-10pt}
\section{Data Source: World of Code}\label{s:sources}

The World of Code~\cite{woc19}(WoC) infrastructure prototype was created
to support the development of theoretical, computational, and statistical
frameworks that discover, collect, and process FLOSS operational
data and construct FLOSS supply chains (SC), identify and quantify
its risks, and discover and construct effective risk mitigation
practices and tools. That prototype stores the huge and rapidly
growing amount of data in the entire FLOSS ecosystem and provides
basic capabilities to efficiently extract and analyze the data at
that scale. WoC's primary focus is on types of analyses that require
global reach across FLOSS projects.
In a nutshell, WoC is a software analysis pipeline starting from the
discovery and retrieval of data, data storage and regular updates, and
enablement of the transformations and data augmentations necessary for 
analytic tasks
downstream. In addition to storing objects from all git repositories,
WoC also provides relationships among them.  For the purpose of this
analysis, we only use a single list of author IDs extracted from
almost 2B commits in WoC. 

WoC data is versioned, with the latest version labeled as Q, 
containing 7.2 billion blobs, 1.8 billion commits, 7.6 billion trees, 16 million tags, 116 million projects (distinct repositories), and 38 million distinct author IDs. 
WoC has collected that data during
November and December of 2019. For more information please consult the
WoC website~\footnote{\url{https://bitbucket.org/swsc/overview/src/master/}}. 
% The proposed grouping into the related projects 
% produced 63854266 such clusters with the largest cluster containing 117248 
% repositories (jtleek\_datasharing).

We used all the author IDs extracted
from WoC for this analysis,which were extracted from the commits,
and are represented by a combination of the authors' names and email
addresses, in the following format: \texttt{first-name
  last-name$<$email-address$>$}. E.g. for an author whose first name
is ``John'', last name is ``Doe'', and email address is
``john@me.com'', the corresponding author id in the WoC dataset
would be: ``\texttt{John Doe <john@me.com>}''. In actual commits the
format may not be exactly preserved, and some extreme examples commits contain Author IDs that may be gigabytes long (due to having very long strings as author name/ email). 

\section{Methods of linking author identities}\label{s:link}
%  \subsection{Description of the Shared Data}

\begin{figure*}[!t]
\centering
\includegraphics[width=\linewidth]{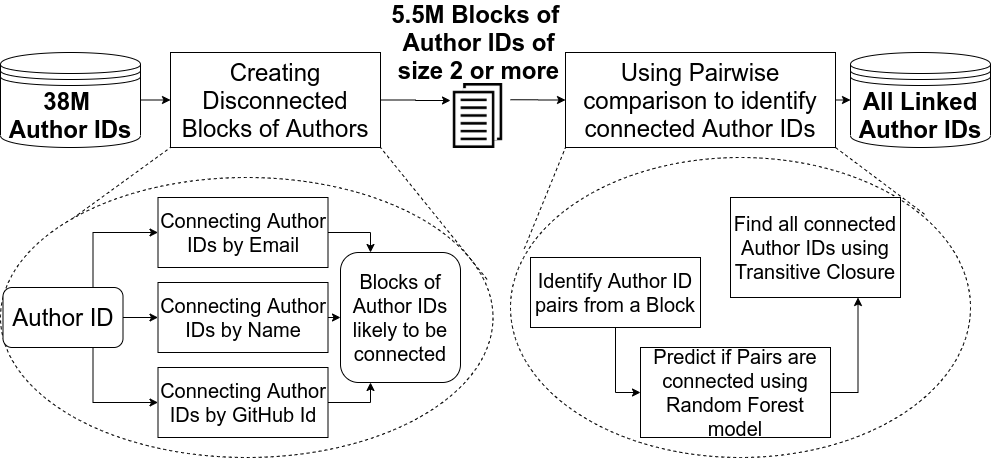}%
\caption{Workflow showing the steps taken for linking author IDs}
\label{fig:flow}
\vspace{-10pt}
\end{figure*}

Studies of online activities such as collaboration in software
development made identity resolution techniques important in the
field of empirical software engineering
research~\cite{german2004mining,german2003automating,Bird:2006:MES:1137983.1138016}.
They help disambiguate identities of people in a software projects
and ecosystems. Data mined from version control systems can be used
for many purposes: \cite{msrData15} builds social diversity
dataset, others measure output~\cite{Gharehyazie2015} and match
developer identity across Open Source Software~\cite{GH_SO}, and/or
link it to mailing lists~\cite{MailingList}. It can also be particularly 
useful in the context of bot detection~\cite{dey2020detecting,dey2020exploratory}
and predicting which pull requests will be merged~\cite{dey2020pull}. Knowing 
the actual number of 
developers who contribute to or use a project can also influence the 
calculations of the number of defects~\cite{dey2018software,dey2020deriving}, or 
the popularity of the project~\cite{dey2018modeling}.

The most accurate methods for identifying connected author
IDs, like ALFAA~\cite{amreen2019alfaa}, require pairwise
comparison of all author IDs, which is infeasible to apply on
millions of author IDs. Therefore, we subdivide the problem by first
creating blocks of author IDs that are likely to be connected, and
then apply the pairwise comparison method for identifying all author
IDs that are actually linked. The workflow used to generate this
data is shown in Figure~\ref{fig:flow}. 

As mentioned earlier, we started the process with all 38M different
author IDs. The idea behind creating the blocks is to maximize the
chance that author IDs belonging to the same group could actually
belong to one developer, while author IDs belonging to different
groups are unlikely to belong to the same developer. We constructed
3 different types of maps for creating the blocks:
\begin{itemize}
    \item \textbf{Author Email to Author ID map:} We created a map
by linking the author IDs that have the same email address, after
checking if the email address is valid using several heuristics.
These heuristics are essential because git commits do not enforce
any strict email field policy, allowing users or their system to use
any arbitrary string as their ``email''. This results in millions of
author ids that contain very common and uninformative emails, such
as \texttt{John<john@example.com>}. Including emails of this type
would result in groups that do not actually represent one
developer. Therefore, to avoid including such email addresses,
we use frequency analysis to determine the
top ``junk'' emails that exist in the author IDs dataset and remove
them. We filter these emails further through other heuristics such
as email length and regular expression matching (the details are in
the provided source code). The map is then generated by sweeping
through this cleaner set of author IDs and grouping all of the IDs
that share the same exact email field. The mapping is presented in
the form of \texttt{email;author\_ID1;author\_ID2;...} The idea is
that author IDs that share the exact same valid email address are
more likely to belong together.
    \item \textbf{Author Name to Author ID map:} We created another
map by linking the author IDs based on author names, using a
similar approach to the author emails. This approach looks only at
the first and last names of the author ID. Similar to the email
field, the name field has just as much flexibility, allowing users
to assign whatever they choose as their first and last name. As a
result, a set of heuristics is applied on the name field to filter
out short and uninformative names. However, we are much more strict
with these heuristics because names are much more difficult to tell
apart than emails (e.g. ``John Doe'' vs. ``Johnathan Doe''). For
example, after filtering out certain names, we also grouped together
similar names and only included them in the final mapping if their
group size was at most 12 author IDs (discussed in-depth
below). This ensured that valid names that passed the original
filter, such as ``John Smith'', were ultimately thrown out to avoid
one developer being responsible for all author IDs containing a
common name. Similar to the approach above, the idea of this mapping
is that author IDs that share the exact same, uncommon name are more
likely to belong together.
    \item \textbf{Author's GitHub handle to Author ID map:} Although
the WoC data has author IDs connected to all git repositories, a
large portion of authors use GitHub. Therefore, we decided to
construct a map between author IDs and the GitHub handles of the
authors, where available, using information from
GHTorrent~\footnote{\url{http://ghtorrent.org/}}. 
\end{itemize}

Finally, we used transitive closure on the first and last maps, along with the map for the uncommon author names to create blocks of author IDs. 
We provide the script to generate the  blocks for all the three map combinations, and the data with the author blocks we generated, with the author IDs and the frequency of the first and last names, and we leave it to the community to decide if it is appropriate to run the second step (name based mapping) of the algorithm on all names for identity resolution.
At the end of the process, we were left with 5,508,119 blocks of size 2 or more, containing 15,177,184 author IDs in total, with the  maximum block size being 42.  

After creating these blocks, we used pairwise comparison to identify
which author IDs are actually linked together. In this way, we
reduced the number of required pairwise comparison from $38M *
38M \approx 1.4*10^{15}$ to maximum $5.5M*42*42 \approx 9*10^9$.
Moreover, since these blocks are independent by assumption, we can
run these tasks in parallel, further reducing the computation time.  

For the pairwise comparison, we first created the author ID pairs by
comparing the Jaro-Winkler distance between the authors' first names,
last names, full names, usernames (from username@domain of email address), 
and the complete email address.
For the model comparing the names, we also used a switched version of
first and last name, since sometimes the authors might
write their last name before their first name. Following this step,
we used a Random Forest model that was used in the implementation of
ALFAA~\cite{amreen2019alfaa}, and was trained using the OpenStack
data shared in that work. If all the author IDs in a block are predicted to be
connected, then we simply use that whole block as a group of IDs
that belong to one developer; otherwise, we identify the pairs that
are identified to be connected, and use transitive closure to create
the final group(s) of author IDs that belong to individual
developer(s) within the block.  

By using the above-mentioned process, we were able to identify that 14,861,538 author IDs, out of the initial 15,177,184 IDs in all the blocks, were linked to at least one other author ID, and we found that these 14,861,538 author IDs belong to 5,427,024 different developers, with the median number of aliases used by an individual developer being 2. 

To evaluate the performance of our method, we enlisted the help of 44 different developers, who identified all the different author IDs that belong to them, and they identified a total of 207 different author IDs that belong to them. We ran our method on these 207 author IDs and checked if our method was able to identify each pair of connected author IDs. Our method resulted in a precision of 0.99 and a recall of 0.84 using this test data. 

\subsection{A Few Problematic Cases:}
We have faced a number of challenges while working on this problem, and we are highlighting a few of them here:
\begin{itemize}
    \item In a number of cases, the authors didn't have a first and last name, but had an alias instead, which meant either the first or the last name was empty for those authors, which made the task of matching them more challenging. In the same vane, a number of authors had non-English language names, and we are not sure how well our method, especially the Jaro-Winkler distance measurement, would work for those cases.
    \item A number of authors had blank names, email addresses, or both, which makes the our method ineffective for identity resolution of those authors.
    \item There are a number of author IDs that are shared between multiple people, e.g. organizational IDs or admin IDs, which makes identity resolution extremely difficult in those cases. 
    \item A few author names are very common, which means there could be multiple people with the same name, and the task of identity resolution is very difficult for those authors. 
    
\end{itemize}

\subsection{Description of the Shared Data:}
We created a dataset with all the 14,861,538 author IDs that were found to have an alias, which is a CSV file with `;' as the separator. If an author was found to have 2 different IDs: \textit{I1, I2}, then it is recorded in the file in 2 separate lines, with the lines being \texttt{I1;I1} and \texttt{I1;I2}, i.e. the first column is the group identifier, which is one of the IDs in a group, and the second column contains the different author IDs in separate lines. The restricted access version and the publicly available versions of the data are available at the links shared in Section~\ref{s:intro}. 

We also shared the source code we used for creating the dataset, including the scripts, the pre-trained Random Forest model we used, and a README file describing how to use the scripts, at the link shared in Section~\ref{s:intro}. These models can be loaded into R language and used in the provided workflow. 

The last part of data we share is the blocking done using all three heuristics and is in a gzipped semicolon separated text file
containing block id, frequency of the first name (number of times it appears in the 38M WoC version Q author IDs), frequency of the last name, full name, email, and Author ID. The largest block contains 993 Author IDs. 
The restricted access and publicly available versions of the data are available at the links shared in Section~\ref{s:intro}.

\vspace{-10pt}
\section{Future Work}\label{s:future}

We plan to (or hope someone from the MSR community would) further refine our heuristics to create the author ID blocks, and potentially increase the block sizes to ensure no potential matches are in different blocks.

We hope to improve upon the models we have now in three ways. First, train them on the larger training sample. Second, add information about name frequencies (see blocking description above) as there were found to be important in prior work~\cite{amreen2019alfaa} and other so called ``behavioural fingerprints''. Finally, we would like to address the issue of homonyms, i.e., Author IDs used by multiple developers by assigning authorship at the commit level.  

\vspace{-10pt}
\section{Limitations}\label{s:lim}

We utilize WoC data collection with all associated limitations of
using that repository and described there~\cite{woc19}. 

Our primary assumption while applying this method is that author IDs in different blocks do not match, which is not always true, but we sacrifice a little accuracy for heavily improved performance by using this method.

The task of identity resolution solely based on name and email ID is very difficult for the problematic cases, as described earlier. Looking at the activity of those authors can be helpful for identity resolution for such cases. 

The accuracy of our approach is not easy to establish, since there is no sizeable Golden dataset that we can use for reference. We tried to circumvent this problem by enlisting the help of 44 developers who identified 207 IDs belonging to them. Still, the size of this test sample is very small, and might not be representative of the whole population.

\vspace{-10pt}
\section{Conclusion}\label{s:conc}

We expect this dataset to spur research that relies on accurate author identities by eliminating the painstaking manual verification that is not even possible for large collections. 
We also expect to update the data as the information from more projects is collected and as larger training datasets and more accurate blocking and matching models are established.

\vspace{-10pt}
\section*{Acknowledgement}
The work has been partially supported by the following NSF awards:
CNS-1925615, IIS-1633437, and IIS-1901102.

\balance
\bibliographystyle{ACM-Reference-Format}
\bibliography{references}

\end{document}